\documentclass[conference,12pt,onecolumn]{IEEEtran}
\normalsize
\usepackage{amsmath, amsthm, amssymb}
\usepackage{siunitx}
\usepackage{filecontents}
\usepackage[numbers,sort&compress]{natbib}
\usepackage{epsfig}
\usepackage{latexsym}
\usepackage{epstopdf}
\usepackage{multirow}
\usepackage{graphicx}
\usepackage[caption=false]{subfig}
\usepackage{algorithm}
\usepackage{algpseudocode}
\usepackage{subfig}
\usepackage{float}

\begin{document}
\title{Energy-Efficient Resource Allocation for Mobile Edge Computing-Based Augmented Reality Applications}

\author{Ali Al-Shuwaili and Osvaldo Simeone }
\maketitle

\begin{abstract}

 Mobile  edge computing is a provisioning solution to enable Augmented Reality (AR) applications on mobile devices. AR mobile applications have inherent collaborative properties in terms of data collection in the uplink, computing at the edge, and data delivery in the downlink.
In this letter, these features are leveraged to propose a novel resource allocation approach  over both communication and computation resources. The approach, implemented via Successive Convex Approximation (SCA), is seen to yield  considerable gains in  mobile energy consumption as compared to conventional  independent offloading across users.

\end{abstract}

\let\thefootnote\relax\footnotetext{A. Al-Shuwaili and O. Simeone are with the Center for Wireless Information Processing (CWiP),  Department of Electrical and Computer Engineering, New Jersey Institute of Technology, Newark, NJ 07102 USA (e-mail: ana24@njit.edu, and osvaldo.simeone@njit.edu).}


\section{Introduction}
 Augmented Reality (AR) mobile applications are gaining  increasing attention due to the their ability to combine computer-generated data with the physical reality. AR applications are computational-intensive and delay-sensitive, and their execution on mobile devices is generally prohibitive when satisfying users' expectations in terms of battery lifetime \cite{van2010survey,liu2013gearing, verbelen2013leveraging}. To address this problem, it has been proposed to leverage mobile edge computing \cite{verbelen2013leveraging, bohez2013mobile, chung2015adaptive, satyanarayanan2009case}. Accordingly, users can offload the execution of the most time- and energy-consuming computations of AR applications to cloudlet servers via wireless access points.

The stringent delay requirements pose significant challenges to the offloading of AR application via mobile edge computing \cite{ verbelen2013leveraging, satyanarayanan2009case}. A recent line of work has demonstrated that it is possible to significantly reduce mobile energy consumption under latency constraints by performing a joint optimization of the allocation of communication and computational resources \cite{ali2016, sar2014}. These investigations apply to generic applications run independently by different users. However, AR applications have the unique property that the  applications of different users share part of the computational tasks and of the input and output data \cite{verbelen2013leveraging,bohez2013mobile }. In this paper, we propose to leverage this property to reduce communication and computation overhead via the joint optimization of communication and computational resources.

 To illustrate the problem at hand, consider the class of AR applications that superimpose artificial images into the real world through the screen of a mobile device. The block diagram of such applications shown in Fig. 1 identifies the following components\cite{verbelen2013leveraging, bohez2013mobile}: (\emph{i}) \textit{Video source}, which obtain raw video frames from the mobile camera; (\emph{ii}) \textit{Tracker}, which tracks the position of the user with respect to the environment; (\emph{iii}) \textit{Mapper}, which builds a model of the environment; (\emph{iv}) \textit{Object recognizer}, which identifies known objects in the environment; and (\emph{v}) \textit{Renderer}, which prepares the processed frames for display. The Video source and Renderer components must be executed locally at the mobile devices, while the most computationally intensive   Tracker, Mapper and Object recognizer components can be offloaded. Moreover, if offloaded, Mapper and Object recognizer can collect inputs from all the users located in the same area, limiting the transmission of redundant information in the uplink across users. Also, the outcome of the Mapper and Object recognizer components can be multicast from the cloud to all co-located users in the downlink.

\begin{figure}[!t] \label{fig1}
        \centering
        \includegraphics[width=0.7\textwidth,height=.5\textheight,keepaspectratio]{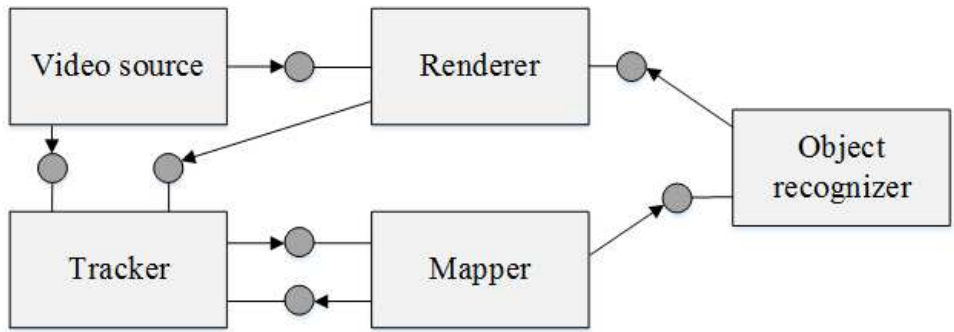}
                \caption{{\small{Example of a component-based model of an AR application \cite {verbelen2013leveraging}. The application includes the Video source and Renderer components,  which need to be executed locally on the mobile device, and the three main components of Mapper, Tracker and Object recognizer, which may be offloaded.}}}
        \label{model}
\end{figure}

In this work,  unlike prior papers \cite{sar2014, ali2016}, we tackle the problem of minimizing the total mobile energy expenditure for offloading under latency constraints over communication and computation parameters by explicitly accounting for the discussed  \textit{collaborative} nature of AR applications. Section II introduces the system model. The  resource allocation problem is formulated  and tackled by means of a proposed Successive Convex Approximation (SCA) \cite{scutari2014distributed} solution in Section III. Numerical results are finally provided in Section IV.

\section {System Model}
We consider the mobile edge computing system illustrated in Fig. 2, in which $K$ users in a set $\mathcal{K}=\{1,\ldots,K\}$ run a  computation-intensive AR application on their single-antenna mobile devices with the  aid  of  a cloudlet  server. The server is attached to a single-antenna Base Station (BS), which serves all the users in the cell using  Time Division Duplex (TDD) over a frequency-flat fading channel. {Following the discussion in Sec. I, we assume that the offloaded application has shared inputs, outputs and computational tasks, which pertain to the Tracker, Mapper and Object recognizer components}.

 To elaborate, let us first review the more conventional set-up, studied in, e.g., \cite{ali2016, sar2014}, in which users perform the offloading of \textit{separate} and independent applications. In this case, offloading for each user $k$ would require: (\emph{i}) \textit{Uplink:} transmitting a number $B_k^I$ input bits from each user $k$ to the cloudlet in the uplink; (\emph{ii}) \textit{Cloudlet processing:} processing the input by executing $V_k$ CPU cycles at the cloudlet; (\emph{iii}) \textit{Downlink:} transmitting $B_k^O$ bits from the cloudlet to each user $k$ in the downlink. {In contrast, as discussed in Sec. I, the collaborative nature of the Tracker, Mapper and Object recognizer components (recall Fig. 1) can be leveraged to reduce mobile energy consumption and offloading latency, as detailed next. We note that the non-collaborative components can potentially also be carried out locally if this reduces energy consumption (see, e.g., [7][8]). We leave the study of the optimization of this aspect to future work}.

\begin{figure}[!t] \label{fig2}
        \centering
        \includegraphics[width=0.6\textwidth,height=.5\textheight,keepaspectratio]{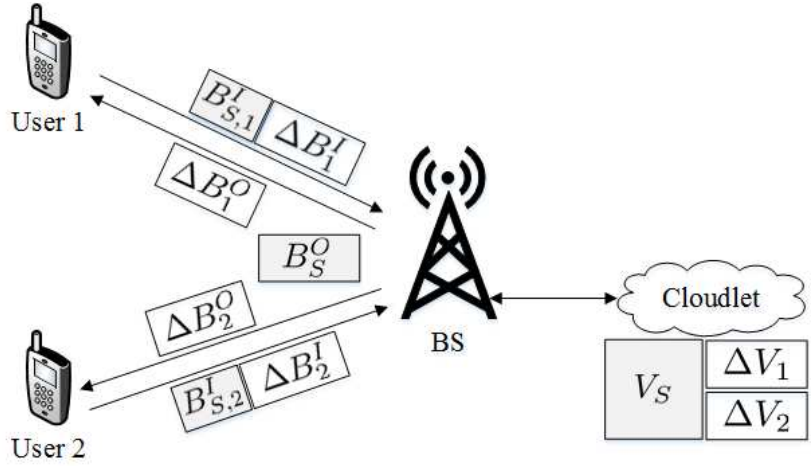}
\caption{{\small{Offloading of an AR application to a cloudlet attached to the BS. Shared data and computations are shaded.}}}
\label{fig:test}
\end{figure}

1) \textit{Shared uplink transmission:} A given subset of  $B_{S}^I \leq \text{min}_k\{B_k^I\}$ input bits is shared among the users   in the sense that it can be sent by \textit{any} of the users to the cloudlet. For example, the input bits to the offloaded Object recognizer component in Fig. 1 can be sent by  {any} of the users in the same area. As a result, each user $k$ transmits a fraction of $B^I_{S,k}$ bits of the $B_S^I$ shared bits, which can be optimized under the constraint $\sum_k B_{S,k}^I = B_S^I$, as well as $\Delta B_k^I=B_k^I-B_{S,k}^I$ bits that need to be uploaded exclusively by user $k$.

2) \textit{Shared cloudlet processing:} Part of the computational effort of the cloudlet is spent producing output bits of interest to all users. An example is the computational task of updating the model of the environment carried out by the mobiles. Therefore, we assume that $V_S \leq \text{min}_k \{ V_k \}$ CPU cycles are shared, whereas  $\Delta V_k=V_k-V_S$ CPU cycles are to be executed for each user $k$.

3) \textit{Multicast downlink transmission:} Some of the output bits need to be delivered to all users. For example, a co-located group of users may need the output bits from the Mapper component for a map update. To model this, we assume that a subset of $B_{S}^O \leq \text{min}_k\{B_k^O\}$ output bits can be transmitted in multicast mode to all users in the cell, while $\Delta B_k^O=B_k^O-B_{S}^O$ bits need to be transmitted  to each user in a unicast manner.

The frame structure is detailed in Fig. 3. Not shown are the training sequences  sent by the users prior to the start of the data transmission frame, which enable the BS to estimate the uplink Channel State Information (CSI), and hence also the downlink CSI due to reciprocity. The CSI is assumed to remain constant for the frame duration. As seen in Fig. 3, in the data frame, the shared communication and computation tasks are carried out first,  followed by the conventional separate offloading tasks, as discussed next.

%
%

\begin{figure}[!t] \label{fig3}
        \centering
        \includegraphics[width=0.8\textwidth,height=.5\textheight,keepaspectratio]{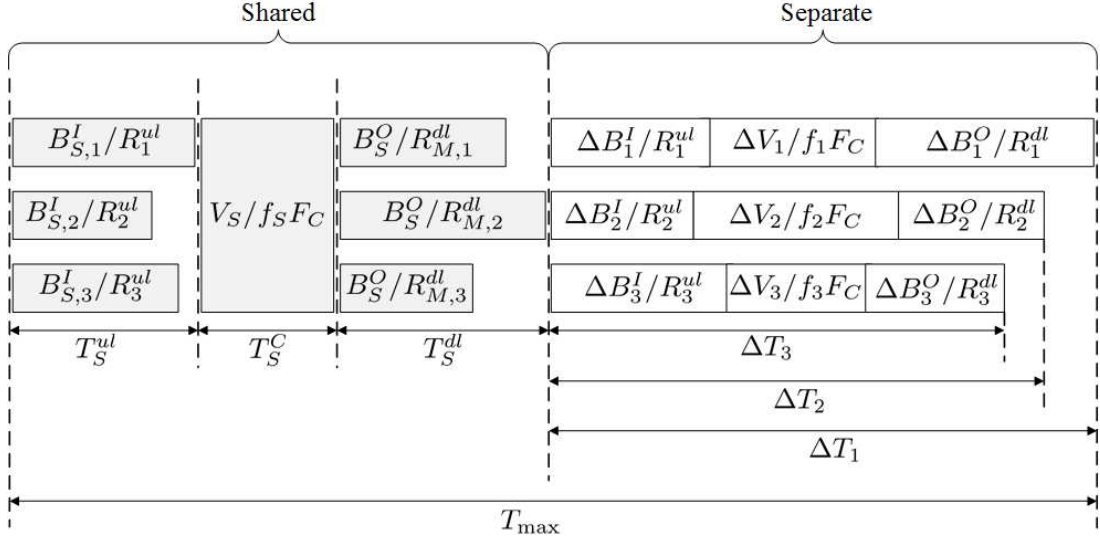}
        \caption{{\small{Data frame structure for a system with $K=3$ users. The preamble containing training sequences is not shown.}}}
        \label{model}
\end{figure}

1) \textit{Uplink transmission:} The achievable rate, in bits/s,  for transmitting the input bits of user $k$ in the uplink is given by
\begin{equation}
R^{ul}_k(P^{ul}_k)=\frac{W^{ul}}{K} \log_2\left(1+\frac{\gamma_k P^{ul}_k}{N_0W^{ul}/K}\right),
\end{equation}
where $P^{ul}_k$ is the transmit power of the mobile device of user $k$; the uplink bandwidth $W^{ul}$ is equally divided among the $K$ users, e.g., using OFDMA; $\gamma_k$ is the uplink and downlink channel power gains of user $k$; and $N_0$ is the noise power spectral density at the receiver. Referring to Fig. 3 for an illustration,  the time, in seconds, necessary to complete the shared uplink transmissions is defined as  $T_S^{ul}=\text{max}_k B_{S,k}^I/R^{ul}_k(P^{ul}_k)$, whereas the time needed for user $k$ to transmit the separate $\Delta B_k^I$ bits is $\Delta B_k^I /R^{ul}_k(P^{ul}_k) $.
The corresponding mobile energy consumption due to uplink transmission is
\begin{equation}
E_{k}^{ul}(P^{ul}_k, B_{S,k}^I) = \left(\frac{{P^{ul}_k}}{R^{ul}_k(P^{ul}_k)}  + l_k^{ul}\right) \left(B_{S,k}^I + \Delta B_{k}^I \right),
\end{equation}
where $l_k^{ul}$ is a parameter that indicates the amount of energy spent by the mobile device to extract each bit of offloaded data from the video source. In (2) and in subsequent equations, we make explicit the dependence on variables to be optimized.

2) \textit{Cloudlet processing:} Let $F_{C}$ be the capacity of the cloudlet server in terms of number of CPU cycles per second. Also, let $f_{k}\geq 0$ and $f_{S}\geq 0$ be the fractions, to be optimized, of the processing power $F_{C}$ assigned to run the  $\Delta V_k$ CPU cycles exclusively for user $k$ and the $V_S$ shared CPU cycles, respectively, so that $\sum_{{k}\in\mathcal{K}} f_{k}\leq 1$ and $ f_S\leq 1$. As shown in Fig. 3, the execution time for the shared CPU cycles  is $T_S^C=V_{S}/(f_{S}F_{C})$ and the time needed to execute $\Delta V_{k}$ CPU cycles of interest to user $k$ remotely is $\Delta V_{k}/(f_{k}F_{C})$.

3) \textit{Downlink transmission:} The common output bits $B^O_{S}$ are multicast to all users. Let $P_M^{dl}$ be the transmit power for multicasting, which is subject to the optimization. The resulting achievable downlink rate for user $k$ is given by
\begin{equation}
R_{M,k}^{dl}(P^{dl}_{M})=W^{dl} \log_2\left(1+\frac{\gamma_k P^{dl}_{M}}{N_0W^{dl}}\right),
\end{equation}
with $R_{M}^{dl}(P^{dl}_{M})=\text{min}_k R_{M,k}^{dl}(P^{dl}_{M})$, with $W^{dl}$ being the downlink bandwidth. The  downlink transmission time to multicast $B_S^O$ bits can hence be computed as $T _{S}^{dl} =  B_{S}^O/R_{M}^{dl}(P^{dl}_M)$ (see Fig. 3).
The $\Delta B_k^O$ output bits intended exclusively for each user $k$ are sent in a unicast manner in downlink using an equal bandwidth allocation, with  rate
\begin{equation}
R_{k}^{dl}(P^{dl}_k)=\frac{W^{dl}}{K} \log_2\left(1+\frac{\gamma_k P^{dl}_k}{N_0W^{dl}/K}\right),
\end{equation}
where $P^{dl}_k$ is the BS transmit power allocated to serve user $k$. The overall downlink mobile energy consumption for user $k$ is
\begin{equation}
         E _{k}^{dl}(P^{dl}_k, P_M^{dl}) = \left(\frac{\Delta B_{k}^O}{R_{k}^{dl}(P^{dl}_k)} + \frac{ B_{S}^O}{R_{M,k}^{dl}(P^{dl}_M)}\right) l_k^{dl},
\end{equation}
where $l_k^{dl}$ is a parameter that captures the mobile receiving energy expenditure per second in the downlink. {We note that we leave the optimization of the bandwidth allocation across users for uplink
and downlink to future work}.

\section{ ENERGY-EFFICIENT RESOURCE ALLOCATION}
In this section, we tackle the minimization of the mobile sum-energy  required for offloading across all users under latency and power constraints. Stated in mathematical terms, we consider the following problem:
\begin{equation*}
\begin{array}{l}
\begin{array}{*{20}{l}}
{\mathop {{\mathop{\rm min}\nolimits} }\limits_{  \bf{z} } }&  \sum\limits_{{k} \in {\cal K}}  {{E_{{k}}^{ul}}}(P^{ul}_k, B^I_{S,k}) + E _{k}^{dl}(P^{dl}_k, P_M^{dl}) 
\end{array}\\
\begin{array}{*{20}{l}}
 \; \text{s.t.}&{{\bf{C}}.\mathbf{1}}& \frac{{ \Delta B_{{k}}^I} }{{R_{{k}}^{ul}(P^{ul}_k)}} +  \frac{\Delta V_k}{f_kF_{C}} + \frac{V_S}{f_SF_{C}}  +\frac{{ \Delta B_{{k}}^O}}{{R_{{k}}^{dl}(P^{dl}_k)}} \le {T_{\text{max}}}-{T_{\text{S}}^{ul}}-{T_{\text{S}}^{dl}} , \forall k \in {\mathcal {K}},  \\
  {}&{{\bf{C}}{\bf{.2}}}& {  \frac{{B_{S,k}^I } }{{R_{{k}}^{ul}(P^{ul}_k)}} \le {T_{\text{S}}^{ul}}} , \forall k \in {\mathcal {K}},\\
{}&{{\bf{C}}{\bf{.3}}}& {  { \frac{{B_{{S}}^O}}{{R_{M,k}^{dl}(P^{dl}_{M})}} } \le {{T}_{\text{S}}^{dl}}} , \forall k \in {\mathcal {K}},\\

{}&{{\bf{C}}{\bf{.4}}}&{\sum\limits_{{k} \in {\cal K}} {{f_{{k}}}}  \leq 1};   0 \le f_S \leq 1;f_{{k}}\ge 0 , \forall k \in \cal{K},  \;\;\;\;\;\;\;\;\;\;\;\;\;\;\;\;\;\;\;\; \\
{}&{{\bf{C}}{\bf{.5}}}&{\sum\limits_{{k} \in {\mathcal {K}}} {{B^I_{{S,k}}}}  = B^I_S}, \;\;\;\;\;\;\;\;\;\;\;\;\;\;\;\;\;\;\;\; \\
{}&{{\bf{C}}{\bf{.6}}}& \sum\limits_{{k} \in \mathcal{K}} P_{{k}}^{dl} \leq {{P}}_{\text{max}}^{dl}; P_{{M}}^{dl} \leq {{P}}_{\text{max}}^{dl}; P_{{k}}^{ul} \leq {{{P}}_{\text{max}}^{ul}, \forall k \in {\mathcal {K}}.} \\
\end{array}
\end{array}\notag \tag{P.1}\label{P.1}
\end{equation*} 
The optimization variables are collected in  vector $\mathbf{z}\triangleq \big( {{{\bf{P}}^{ul}},{{{\bf{B}}_S^{I}}},{{\bf{f}}},{{\bf{P}}^{dl}}},{{{P}}_M^{dl}}, {{T}_{\text{S}}^{ul}}, {{T}_{\text{S}}^{dl}} \big)$, where $\mathbf{P}^{ul} \triangleq ({{P}^{ul}_{k}})_{k\in\mathcal{K}}$, $ \mathbf{B}_S^I \triangleq (B_{S,k}^I)_{k\in\mathcal{K}} $, $\mathbf{f} \triangleq \left(({{f}}_{k})_{k\in\mathcal{K}}, f_S \right)$, $\mathbf{P}^{dl} \triangleq ({{P}^{dl}_{k}})_{k\in\mathcal{K}}$, and we defined $\mathcal{Z}$ as the feasible  set of problem (P.1).
As illustrated in Fig. 3, constraints C.1-C.3 enforce that the execution time of the offloaded application to be less than or equal to the maximum latency of ${T}_{\text{max}}$ seconds. Constraints C.4-C.5 impose the conservation of computational resources and shared input bits,  and  C.6 enforces transmit power constraints at BS and users.

 Problem (P.1) is not convex because of the non-convexity of the energy  function ${{E_{{k}}^{ul}}}(P^{ul}_k, B^I_{S,k})$ and of the latency constraints C.2, which we can rewrite as $g_k(P^{ul}_k, B^I_{S,k}) \leq T_S^{ul}, \forall k \in \cal{K}$.  We address this issue by developing an SCA solution following \cite{scutari2014distributed}. Theorem 2 in \cite {scutari2014distributed} shows that the SCA algorithm   converges to a stationary point of the { non-convex} problem (\ref{P.1}). Furthermore, such convergence requires a number of iterations proportional to $1/\epsilon$, where $\epsilon$ measures the desired accuracy in terms of the  stationarity metric $\|{F}\left( \mathbf{z} \right) \|_2^2$  defined in  \cite[Eq. (6)]{cannelli2017asynchronous}.

 In order to apply the SCA method, we need to derive convex approximants for the  functions ${{E_{{k}}^{ul}}}(P^{ul}_k, B^I_{S,k})$ and $g_k(P^{ul}_k, B^I_{S,k})$   that satisfy the conditions specified in \cite[Sec. II]{scutari2014distributed}. Using such approximants, we obtain the SCA scheme detailed in  Algorithm 1. In the algorithm, at each iteration $v$, the unique solution  ${\bf{\hat z}}\left( {{\bf{z}}\left( v \right)} \right)\triangleq \big( {{{\bf{\hat P}}^{ul}},{{{\bf{\hat B}}_S^{I}}},{{\bf{\hat f}}},{{\bf{ \hat P}}^{dl}}},{{{ \hat P}}_M^{dl}}, {\hat T}_S^{ul}, \hat T_S^{dl} \big)$ of the following strongly convex problem
\begin{align*}
& {\bf{\hat z}}\left( {{\bf{z}}\left( v \right)} \right) \triangleq \underset{{  {{\bf{z}}}}}{\text{argmin}}
   \begin{aligned}[t]
       \sum_{k \in \mathcal{K}}  & \tilde{E}\left( \mathbf {z}_k;\mathbf{z}_k\left(v\right) \right)
   \end{aligned} \notag \\
    & \begin{aligned}[t]
       & \;\;\;\;\;\;\;\;\;\;\;\;\;\; \text{s.t.} \\       
       \mathbf{C.2} & \;\;\;\; \tilde{g}_{k}\left(P^{ul}_k, B^I_{S,k};P^{ul}_k(v), B^I_{S,k}(v)\right)  \le {T_{\text{S}}^{ul}} , \forall k \in {\mathcal {K}},  \\
   \end{aligned}  \\
   & {} {{\bf{C.1}},\bf{C}}.\mathbf{3}-{\bf{C}}.\mathbf{6} \;\; \text{of (P.1)}, \tag{P.2}\label{P.2}
\end{align*}
 is obtained, where we have defined $\mathbf{z}_k\triangleq \big( P^{ul}_k,B_{S,k}^{I},f_k,f_S,P^{dl}_k,P_M^{dl}, T_S^{ul}, T_S^{dl} \big)$ as well as $ \tilde{E}_k\left( \mathbf {z}_k;\mathbf{z}_k\left(v\right) \right) \triangleq
       \tilde{E}_{k}^{ul}\left( \mathbf {z}_k;\mathbf{z}_k\left(v\right) \right) + E _{k}^{dl}(P^{dl}_k, P_M^{dl})  $. The approximants functions $\tilde E_k^{ul}\left( .;. \right)$ and $\tilde{g}_{k}\left(.;. \right)$ are discussed next.
The approximant  $\tilde{E}_{k}^{ul}\left({\mathbf {z}_k; \mathbf{z}_k}\left(v\right) \right)$ around the current  feasible iterate $\mathbf{z}_k(v)$ can be obtained following \cite [Sec. III, Example \#8]{scutari2014distributed} as
 \begin{equation}
 \begin{split}
&\tilde{E}_{k}^{ul}\left( {\bf {z}}_k;{\bf{z}}_k\left(v\right) \right)  = \;  \frac{{P^{ul}_k(v) \left(B_{S,k}^I(v)+\Delta B_{k}^I\right)  }} {R^{ul}_k\left(P^{ul}_k\right)} \\
+ &  \frac{{P^{ul}_k(v) \left(B_{S,k}^I+\Delta B_{k}^I\right)  }} {R^{ul}_k\left(P^{ul}_k(v)\right)}+\frac{{P^{ul}_k \left(B_{S,k}^I(v)+\Delta B_{k}^I\right)  }} {R^{ul}_k\left(P^{ul}_k(v)\right)}   \\
 + & \bar E_k^{ul}\left(\mathbf{z}_k;\mathbf{z}_k(v)\right) + l_k^{ul} \left(B_{S,k}^I + \Delta B_{k}^I \right) ,
\end{split}
\end{equation} where $ \bar E_k^{ul}\left(\mathbf{z}_k;\mathbf{z}_k(v)\right) \triangleq \left(\mathbf{z}_k-\mathbf{z}_k(v)\right)^T \Psi \left(\mathbf{z}_k-\mathbf{z}_k(v)\right)$, with $\Psi$ being a diagonal matrix with non-negative elements  $\tau_{P^{ul}},\tau_{{B}_S^{I}},\tau_{{f}},\tau_{{f}_{S}},\tau_{{P}^{dl}},\tau_{{P}^{dl}_M},\tau_{{T}^{ul}_S}$  and $\tau_{{T}^{dl}_S}$. For the second approximant, in light of the relation $g(x_1,x_2) = x_1 x_2 = 1/2 (x_1+x_2)^2 - 1/2 (x_1^2 + x_2^2)$, a convex upper bound is obtained as requested by SCA by linearizing the concave part of $g_k(P^{ul}_k, B^I_{S,k})$\cite[Sec. III, Example \#4]{scutari2014distributed}, which results in
\begin{equation}
\begin{split}
   &\tilde{g}_{k}\left(P^{ul}_k, B^I_{S,k};P^{ul}_k(v), B^I_{S,k}(v)\right) =    \; \frac{1}{2} \Bigg(  \left(B^I_{S,k} + \frac{1}{R_{{k}}^{ul}(P^{ul}_k)}\right)^2 \\
   & - \left(B^{{I}}_{S,k}(v) \right)^2 - \left(\frac{1}{R_{{k}}^{ul}(P^{ul}_k(v))^2} \right) \Bigg) - \Bigg( (B^I_{S,k}(v) \left(B^I_{S,k}-B^I_{S,k}(v)\right) \\
   & - \frac{R_{{k}}^{ul}\left(P^{ul}_k(v)\right)}{\left(1+\frac{\gamma_k P^{ul}_k(v)}{N_0W^{ul}/K}\right)R_{{k}}^{ul}\left(P^{ul}_k(v)\right)^4}
    \bigg(\frac{1}{R_{{k}}^{ul}\left(P^{ul}_k\right)} - \frac{1}{R_{{k}}^{ul}\left(P^{ul}_k(v)\right)}\bigg) \Bigg).
   \end{split} 
  \end{equation}
The convexity of (7) is established by noting that the second term in the right-hand side is the reciprocal of the rate function (concave and positive) and the fourth power (convex and non-decreasing) of a convex  function is convex \cite{boyd2004convex}.

\begin{algorithm}
\caption{SCA Algorithm }\label{alg:SCA_Alg}
\begin{algorithmic}[1]
    \State \textit{Initialization}: ${\bf{z}}\left( 0 \right) \in {\cal Z}$; $\alpha  = 10 ^{-5}$; $\epsilon  = 10^{-5}$; $v = 0$; $\tau_{P^{ul}},\tau_{{B}_S^{I}},\tau_{{f}},\tau_{{f}_{S}},\tau_{{P}^{dl}},\tau_{{P}^{dl}_M},\tau_{{T}^{ul}_S},  \tau_{{T}^{dl}_S} > 0$.
        \State Compute ${\bf{\hat z}}\left( {{\bf{z}}\left( v \right)} \right)$ from (\ref{P.2}).
    \State Set ${\bf{z}}\left( {v + 1} \right) = {\bf{z}}\left( v \right) + \delta \left( v \right)\left( {{\bf{\hat z}}\left( {{\bf{z}}\left( v \right)} \right) - {\bf{z}}\left( v \right)} \right)$, with $\delta \left( v \right) = \delta \left( {v - 1} \right)\left( {1 - \alpha \delta \left( {v - 1} \right)} \right)$.
    \State If $\|{F}\left( \mathbf{z}\left(v\right) \right) \|_2^2 \leq \epsilon$, stop.
    \State Otherwise, set $v \leftarrow v + 1$, and return to step $2$.
\end{algorithmic}
\end{algorithm}

\section{Numerical Results}
In this section, we provide numerical examples with the aim of illustrating the advantages that can be accrued by leveraging the collaborative nature of AR applications for mobile edge computing. We consider a scenario where eight users are randomly deployed in a small cell. The radio channels are Rayleigh fading and the path loss coefficient is obtained based on the small-cell model in \cite{3gpp2010} for a carrier frequency of 2 GHz. The users' distances to the BS are randomly uniformly selected between 100 and 1000 meters and the results are averaged over multiple independent drops of users' location and of the fading channels. The noise power spectral density is set to $N_0=-147$ dBm/Hz. The uplink and downlink  bandwidth is $W^{ul} = W^{dl} = 10$ MHz. The uplink and downlink power budgets are constrained to the values ${P}_{\text{max}}^{ul}=50$ and ${P}_{\text{max}}^{dl}=60$ dBm, respectively. The cloudlet server processing capacity is $F_C=10^{10}$ CPU cycles/s \cite{verbelen2013leveraging}. We also set $l_k^{ul}=1.78\times10^{-6}$ J/bit \cite{7471353}, $l_k^{dl}=0.625$ J/s \cite{d2009} and $\epsilon=10^{-5}$.

 The size of the input data generally depends on the number and size of the features of the video sources obtained  by the mobiles that are to be processed at the cloudlet. Here we select  $B_k^I= 1$ Mbits, which may correspond to the transmission of $ 1024 \times 768$ images \cite{chung2015adaptive}.
 A fraction of the input bits $B_S^I=\eta B_k^I$ bits can be transmitted cooperatively by all users for some parameter $0 \leq \eta \leq 1$. The required CPU cycles of the offloaded components is set to $V_k = 2640 \times B_k^I$ CPU cycles, representing a computational intensive task \cite{Miettinen2010EE}. The shared CPU cycles are assumed to be $V_S=\eta V_k$ for the same sharing factor $\eta$. The output bits are assumed to equal  the amount of input bits $B^O_k = B^I_k = 1$ Mbits with shared fraction $B_S^O=\eta B_k^O$. Practical latency constraints for AR applications are of the order of $ 0.01$ s \cite{chung2015adaptive, bohez2013mobile}. {Throughout our experiments, we found that the accuracy $\epsilon = 10^{-5}$ was obtained within no more than $25$ iterations}.




------------------------------------------------------------------------------------------------
\begin{figure}[!t] \label{fig4}
        \centering
        \includegraphics[width=0.6\textwidth,height=.5\textheight,keepaspectratio]{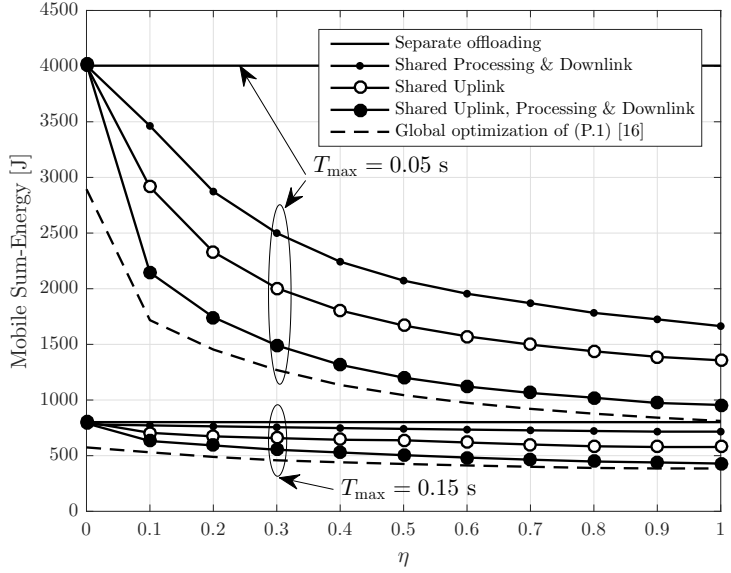}
        \caption{{\small{Average  mobile sum-energy consumption versus the fraction $\eta$ of shared data in uplink and downlink and of shared CPU cycles executed at the cloudlet.}}}
        \label{model}
\end{figure}

For reference, we compare the performance of the proposed scheme, in which uplink and downlink transmissions and cloudlet computations  are shared as described, with the following offloading solutions: \emph{(i) Shared Cloudlet Processing and  Downlink Transmission}: CPU cycles and output data  are shared as described in Sec. II,  while the input bits $B_k^I$ are transmitted by each user individually, i.e., we set $B_{S,k}^I=0$ and $\Delta B_k^I = B_k^I$ for all $k \in \mathcal{K}$; and  \emph{(ii) Shared Uplink Transmission}: Only the input bits  are shared as discussed, while  no sharing of computation and downlink transmission takes place, i.e., $\Delta V_k = V_k$ and $\Delta B_k^O = B_k^O$ for all $k \in \mathcal{K}$. {We also include for reference the result obtained by solving problem (P.1) using the global optimization BARON software running on NEOS server with a global optimality tolerance of $10^{-6}$ \cite{bar}}.  As shown in Fig. 4, for $T_{\text{max}} = 0.05$ s and $\eta=0.3$, the \emph{ Shared Cloudlet Processing and  Downlink Transmission} scheme achieves energy saving   about 37\%  compared to separate offloading (which sets $\Delta B_k^I = B_k^I$, $\Delta V_k = V_k$ and $\Delta B_k^O = B_k^O$ for all $k \in \mathcal{K}$). This gain can be attributed to  the increased time available for uplink transmission due to the shorter  execution and downlink transmission periods, which reduces the associated offloading energy. Under the same conditions, the energy saving of around 50\%  with respect to separate offloading brought by \emph{ Shared Uplink Transmission}   is due to the ability of the system to adjust the fractions of shared data  transmitted  by each user in the uplink based on the current channel conditions. These two gains combine to yield the energy saving of the proposed shared data  offloading scheme with respect to the conventional separate offloading of around 63\%. Both separate and shared offloading schemes have similar energy performance for the relaxed delay requirement of $T_{\text{max}}=0.15$ s, which can be met with minimal mobile energy expenditure even without sharing communication and computation resources. {The figure also shows that SCA yields a solution that is close to the global optimum, for this example}.

\bibliographystyle{ieeetran}
\bibliography{ww}

\end {document}